# A Novel Twitter Sentiment Analysis Model with Baseline Correlation for Financial Market Prediction with Improved Efficiency

Xinyi Guo
*Department of Digital Humanities*
*King's College London*
London, UK

Jinfeng Li*
*Department of Electrical and Electronic Engineering*
*Imperial College London*
London, UK
*jinfeng.li@imperial.ac.uk

***Abstract*— A novel social networks sentiment analysis model is proposed based on Twitter sentiment score (TSS) for real-time prediction of the future stock market price FTSE 100, as compared with conventional econometric models of investor sentiment based on closed-end fund discount (CEFD). The proposed TSS model features a new baseline correlation approach, which not only exhibits a decent prediction accuracy, but also reduces the computation burden and enables a fast decision making without the knowledge of historical data. Polynomial regression, classification modelling and lexicon-based sentiment analysis are performed using R. The obtained TSS predicts the future stock market trend in advance by 15 time samples (30 working hours) with an accuracy of 67.22% using the proposed baseline criterion without referring to historical TSS or market data. Specifically, TSS's prediction performance of an upward market is found far better than that of a downward market. Under the logistic regression and linear discriminant analysis, the accuracy of TSS in predicting the upward trend of the future market achieves 97.87%.**

## I. Introduction

Delivery of an accurate and efficient social networks sentiment analysis model targeting financial market prediction is of research and development interest with the prevalence of Internet of Things (IOT) in the 5G era, which massively changed the way that people interact with each other, and also the way that social media content is generated. Online social media network has become indispensable for public to broadcast their opinions freely and timely. With an ever-increasing numbers of media websites providing Application Programming Interface (API) and database access to general public, user-generated content (UGC) is arguably a big data that reflects the true public sentiment, as compared with that by traditional face-to-face surveys, in which case interviewees may be nervous to express their opinions genuinely. Social networks sentiment analysis has thereby become a popular research topic targeting decision making in different industries, in particular attracting an increasing attention in the financial market. Bollen [1] first used a Twitter mood indicator as an investor sentiment index based on text-mining technologies and predicted the stock price with an accuracy of 87% in their case study, which kicks off the ever-increasing research interest in the media sentiment analysis for financial market prediction. Through the last few years, more and more economic and financial scientists have attempted to verify and improve the power of new media sentiment analysis in financial prediction using different models. However, existing sentiment analysis models struggle to meet the more and more stringent requirement for financial market prediction. Few study to our knowledge has centred upon the trade-off between prediction accuracy and efficiency. Current social networks sentiment analysis method (e.g. Twitter sentiment analysis model in this study) and the mainstream econometric method (e.g. closed-end fund discount index) are still fundamentally limited in their dependence on historical data, which compromises the efficiency for decision making.

In this work, we build up a high-performance Twitter sentiment analysis solution targeting the stock market prediction and conduct a comparative study against traditional investor sentiment analysis (ISA) methods. In addition to statistical techniques, the proposed Twitter sentiment score (TSS) model leverages machine learning with state-of-the-art lexicon-based sentiment analysis and text corpus observations. A particular novelty of this work lies in the definition of a TSS baseline as a fast estimation guideline for the stock market trend (only a comparison operation is required), which decouples historical modelling datasets and simplifies data analytics for targeted users or decision makers without the knowledge of historical TSS and market data. Another novelty of the work lies in incorporating corpus analysis with TSS for model benchmarking and verification. Section II reviews the conventional tools for investor sentiment measurement and introduces the emerging media-platform sentiment analysis applied to financial market prediction. Section III outlines the methodology for the proposed Twitter sentiment model setup. Section IV presents testing results of the proposed TSS baseline correlation model, discusses its advantages and limitations, and provides an outlook for potential application opportunities.

## II. Investor Sentiment Analysis Approach

Investor sentiment analysis (ISA) is primarily based on indexes from econometric modelling and the emerging social media sentiment analysis as reviewed below.

### *A. Closed-end Fund Discount (Premium) Method*

A dominating index characterizing the econometric modelling method is the closed-end fund discount (CEFD), the theory of which can be dated back to [2], which claims that the closed-end fund is always traded at a price lower than its face value, i.e. at a discount value. By definition in eq.1, CEFD is a discount ratio given by the difference between net value and traded value divided by the net value.

$$CEFD = \frac{net\ value\ per\ unit - traded\ value\ per\ unit}{net\ value\ per\ unit} \times 100\% \quad (1)$$

The closed-end fund value is at a discount when CEFD > 0, while at a premium if CEFD < 0. As the opposite number of CEFD, closed-end fund premium (CEFP) has a positive correspondence with investor sentiment, i.e. a larger CEFP indicates a more positive investor sentiment, while a smaller CEFP suggests a more negative investor sentiment. The discounted value phenomenon was summarized as a closed-end fund puzzle [3]. Economists from Harvard University [4] [5] offered a complete explanation for the puzzle, i.e. it is the investor sentiment that influences their confidence in transactions and thus affects the asset pricing. By way of illustration, Lee [4] argued that less confident investors in the market would lead to a decreased pricing due to their concern that a high pricing would attract few buyers and consequently their funds are difficult to circulate. The rationale of this theory was further examined and verified by [5]. In the following decades, many economists employed the closed-end fund theory to measure the investor sentiment, e.g. the experiments done by [6]. Until now, CEFD is still widely be employed as a measure of sentiment by economists to deduce investors' potential transaction behaviour (buy or sell) in the financial market.

However, the CEFD method scratches the surface of the complex stock market problem and is fundamentally limited by its heavy dependency on historical transaction data for data mining and sentiment modelling. This method is only rigorously valid under the assumption that market price and data produced by market transactions can fully reflect the market situations. However, such assumption has long been fraught with controversy, as the effect of human nature has not been factored in. Arguably, the investor behavioural nature coming into play could bring a huge level of uncertainty and even tip the balance. In many cases, the market price is no longer a true representation of the market condition. To bridge the gap, scientists from the fields of digital culture has investigated this problem in an interdisciplinary attempt by opinion mining in social networks corresponding to investor sentiment.

### B. Twitter Sentiment Analysis

While econometric indexes are obtained by processing the historical stock market data, social media sentiment analysis relies on mining text data created by public. Among different social networks, Twitter is of high research impact for investor sentiment analysis in the finance regime. As a micro-blog, Twitter allows public users to freely express opinions and feelings, comment on an event, and show their locations on a tweet, which is limited within 280 characters. The tedious expressions free environment facilitates opinion mining for a timely sentiment analysis. Due to the accessibility to large timely updated datasets with condensed expressions, Twitter API has evolved into a key tool for sentiment analysis research, which helps in the formation of the research subject "Twitter sentiment analysis" [7]. Following the pilot research [1] [8] into Twitter sentiment analysis with stock market, other attempts have been made to verify and extend the Twitter sentiment's prediction potential for the financial market. This work connects to Twitter API for acquiring original opinion-mining datasets, and proposes a new data processing method for improving the efficiency without degrading the prediction accuracy as outlined later.

### C. Machine Learning and Lexicon-based Classification

Machine learning trains a sentiment classification model best matching with the datasets in highest accuracy. Work conducted by [9] researched into sentiment analysis by a variety of machine learning techniques, reporting that the supervised learning model by support vector machines (SVM) achieves the highest accuracy of 82.9% in the non-topic-based sentiment text-content classification. [10] attempted machine learning to transfer the text opinion to data accurately by a trained classifier for neutral, positive and negative words and quantify the text sentiment into a score. This method suffers from labelling massive tweets for classification. Following that, [7] proposed an ontology-based sentiment analysis, which still struggles to eliminate the distortion due to implied sentiment meanings in certain sentences.

Lexicon-based method is free from the time-consuming procedure of labelling training sets and the over-fitting problem as encountered in the machining learning sentiment analysis. Having an opinion lexicon is still far from enough for accurate sentiment analysis because of the difficulty in defining "language sentiment" in a certain word context. It remains highly debatable on how to define the "true sentiment" of a sentence expression. By way of illustration, the phrase "read the book" is a positive expression in the review of a book, whereas a negative expression in commenting a movie. Moreover, people sometimes use ironies to express opinions, in which case that a positive word could indicate a negative meaning in the context. By a comparative review between [10] and [11], machine learning approach tends to achieve a slightly higher accuracy in multi-domain sentiment analysis (86.4%) than that by lexicon-based sentiment analysis (75%-80%). In this work, sentiment analysis on Twitter posts with a single domain of topic (i.e. stock) is conducted incorporating a lexicon-based classification method with a predefined corpus to calculate the sentiment score for the text.

## III. Twitter Sentiment Analysis Model Setup

### A. Data Sources and Collection

The indexes of interest in this research are Twitter sentiment score (TSS, to be collected and calculated), Financial Times Stock Exchange 100 Index (FTSE 100), and closed-end fund discount (CEFD) or closed-end fund premium (CEFP). R Studio is employed for data collection, data processing, results presentation and visualisation. R not only has the API to collect data from selected media platform (i.e. Twitter) by "TwitterR", but also has functions to collect the stock time-series data FTSE 100 for modelling and analysis. Twitter sentiment data is collected from June 14$^{th}$ to Aug. 30$^{th}$ in 2017, when the UK stock market opens from 8:00. to 16:00 (without closing at lunch but closed during weekends). Data is collected every two hours (i.e. 8:00, 10:00, 12:00, 14:00 and 16:00). At each time sample unit, R performs text-mining of 5000 tweets. For each tweet, a positive word is attributed a point of +1, while a negative word is assigned –1. Summing up scores from all sentiment words in a tweet amounts to the Twitter sentiment score (TSS) of the tweet. Mean value of all 5000 tweets' TSS in one time sample unit is obtained as the real-time TSS of that time spot. The calculation process of Twitter sentiment score (TSS) is outlined in Fig. 1.

To further understand the stock market's fluctuations and benchmark with the proposed TSS model, we conduct a word corpus analysis of the collected tweets at each time sample by using an R function called "wordcloud", with the produced bigger and bolder characters indicating a higher frequency appearance in the collection of tweets.

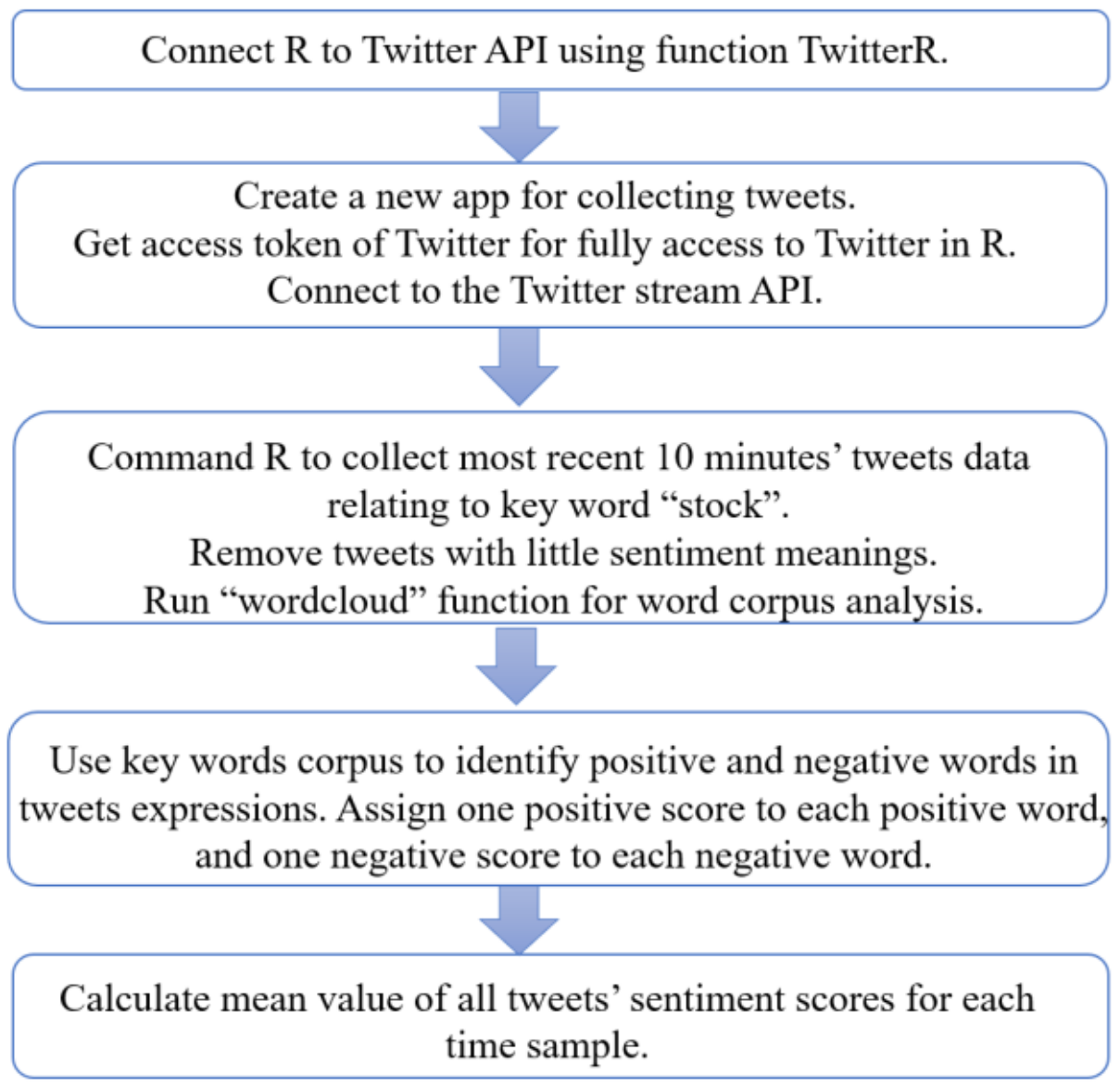

Fig. 1 TSS generation procedure at each time sample.

At each time sample when TSS is computed, data stream of FTSE 100 (representing the stock price) and CEFD are collected from Bloomberg and Investec, respectively, subsequently being imported into R. As given by eq.2, CEFD in this research is calculated from all 655 closed-end funds currently in the UK with a weighted average by their trading volumes, where $i$ denotes a specific closed-end fund, total volume is the sum of all traded closed-end fund volumes.

$$CEFD = \sum_{i=1}^{n=655}(CEFD_i \times \frac{volume_i}{total\ volume}) \qquad (2)$$

## B. Sentiment Model and Data Analytics

Based on the data collected, sentiment modelling combines regression, classification models and lexicon-based sentiment analysis in generating the stock market predictor model. First, we produce the time series scattering plots of the two indicators, i.e. TSS and CEFD (CEFP) compared with FTSE 100 stock price and compute their relative lag (or lead) time periods. Afterwards, a parametric regression model is used for polynomial fitting the lag (or lead) period FTSE 100 with the two indicators separately, and quantifying correlation coefficients for the regression model fitting in search for a predictable time period in their co-vibrations with stock price. Subsequently, the prediction performance of the two indicators versus the lag (or lead) FTSE 100 under regression models is statistically evaluated by quantifying R square, F statistics and p value. Non-parametric classification models, i.e. logistic regression, linear discriminant analysis (LDA), and quadratic discriminant analysis (QDA) are applied for fine tuning of the model-fitting between the predictable time period lag (or lead) FTSE 100 and the two indicators, respectively.

## C. Baseline Correlation for Decoupling Historical Data

Based on the optimised regression model produced, here we propose a new prediction criterion for real-time decision making. The method is based on an intelligent selection of a horizontal baseline with the value $TSS_b$ (constant over time) that serve as a direct indication for market going up or down, e.g. if the latest TSS > $TSS_b$, market will rise; if TSS < $TSS_b$, market will drop. Instead of coupling the historical TSS evolution trend for a decision tree, the proposed baseline approach only requires the current TSS value for market prediction, hence improves the efficiency of decision-making significantly. Prediction accuracy of the approach can be defined as the number ratio of time samples that agree with the aforementioned baseline correlation rule over the total time samples.

# IV. Modelling Results and Analysis

## A. Regression Modelling Results

We perform linear regression of both TSS and CEFP with stock price FTSE 100. The TSS values produced by the proposed procedure are plotted versus a same time series with the FTSE 100. Spacing between adjacent time samples is two working hours. As shown in Fig. 2, the trend of TSS exhibits a high level of similarity with that of the stock price assuming lagging (moving backward) the FTSE 100 curve by a few time samples, i.e. the trend of TSS occurs in advance of FTSE 100. To quantify this trend-time relevance, FTSE 100 is moved backward by certain time samples in the gray arrow direction, i.e. to the left in the graph. By way of illustration in Fig. 3, an updated result of 15-time lag FTSE 100 is presented, which shifts the FTSE 100 curve to the left by 15 time samples and compares with the unchanged TSS time series in trend. Obviously, TSS and the 15-time lag FTSE 100 present a simultaneously vibrating trend largely similar with each other (going up or down at the same time). In recognition for the time-relevance between TSS and FTSE 100, TSS maybe instrumental to predict the stock price in advance by 15 time samples (amounting to 30 working hours).

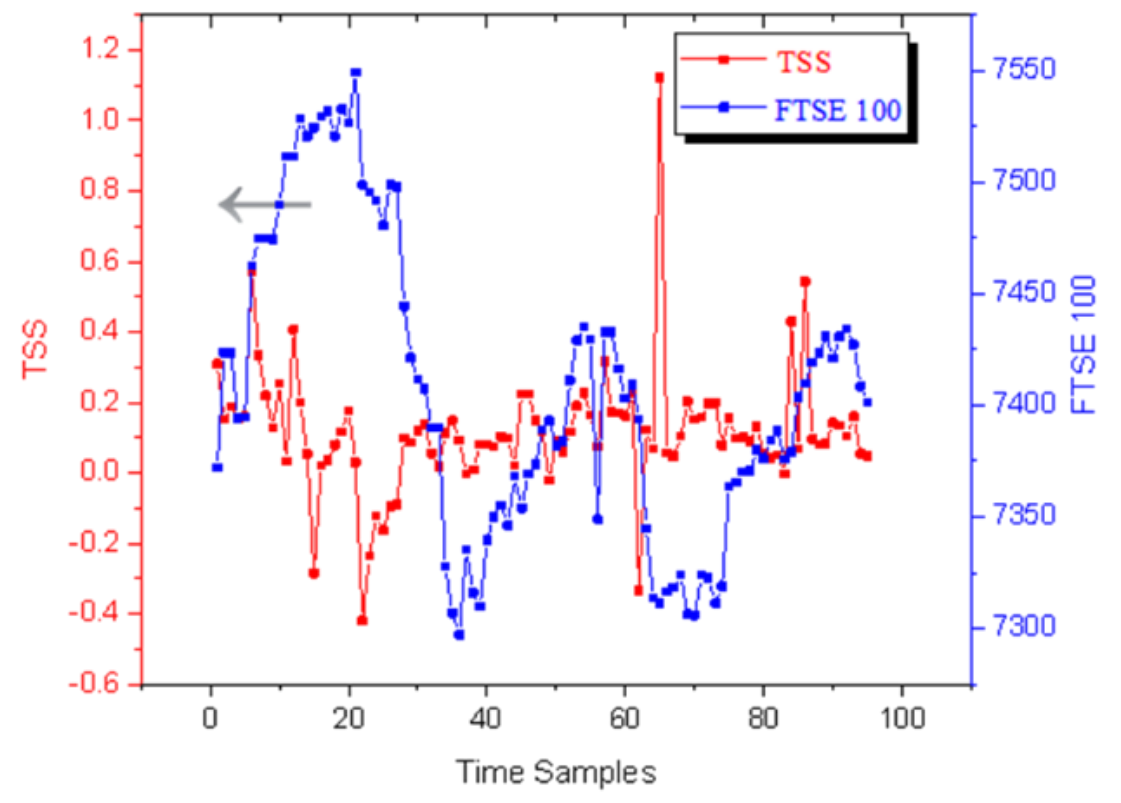

Fig. 2 Results of TSS and FTSE 100 (time sample spacing =2 hours).

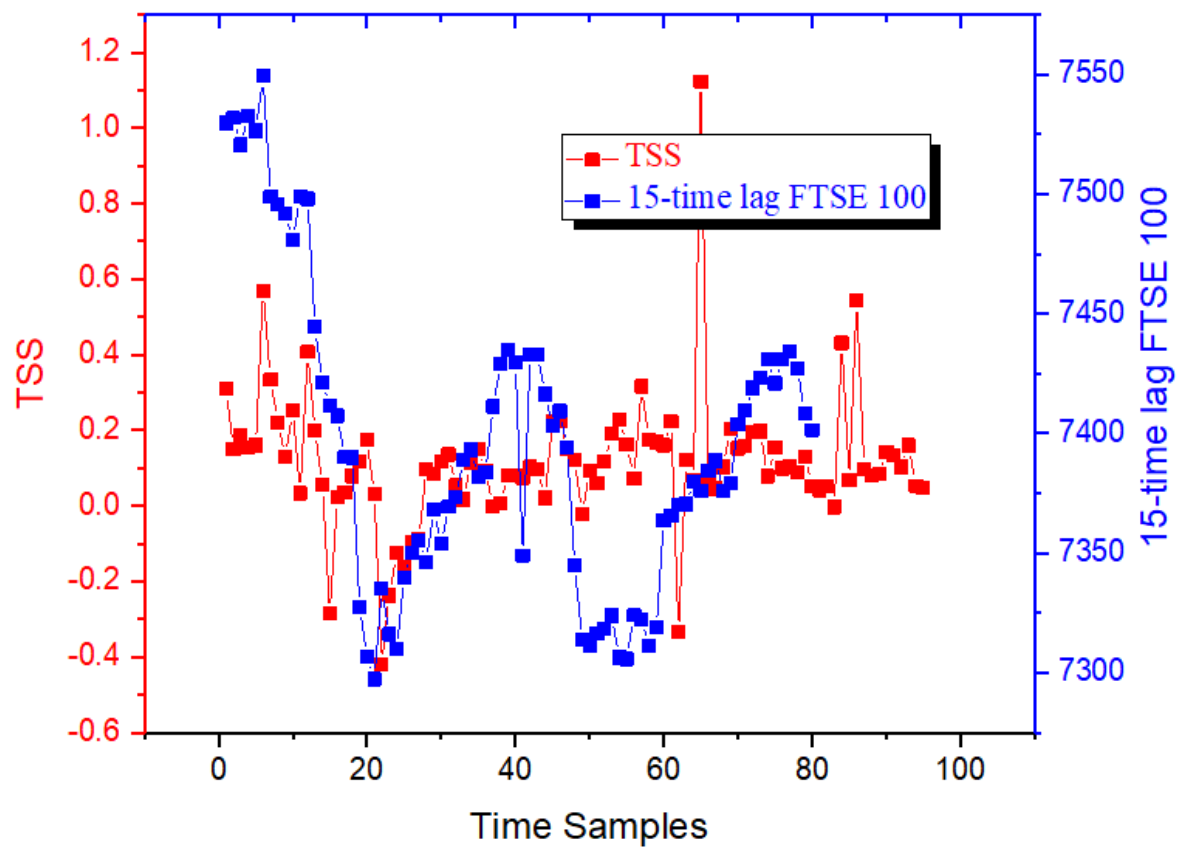

Fig. 3 Results of TSS and 15-time lag FTSE 100.

To mathematically quantify the time-relevance in trends, a polynomial model of 9th order is employed to fit TSS and the 15-time lag FTSE 100, as depicted in Fig. 4. R-square values are obtained as 0.22 and 0.91, respectively, i.e. with the time going, the fitted polynomial model could effectively account for 22% and 91% variation of TSS and the 15-time lag FTSE 100, respectively. The F-statistic in this polynomial modelling is significantly different from zero and p value is far smaller than 5% under the confidence interval of 95%, which proves the effectiveness of the fitting.

To precisely quantify the level of coexistence of the observed simultaneously-varying trend between TSS and the 15-time lag FTSE 100, we strategically mark six rectangular regions and fill them with different colors (i.e. four regions in light blue and the other two in light yellow) based on a quantitative comparison of the TSS value against a base line of $TSS_b$ =0.14453 we choose, as well as an evaluation of TSS's prediction capability on the stock price (capable or incapable). For most of the rectangular regions, we observe that the stock price FTSE 100 is either increasing if the TSS value is higher than the baseline value, or decreasing if the TSS value is lower than the baseline value. We mark these regions with this characteristic in light blue, indicating that the TSS value and its baseline can tell the trend of stock market in advance by 15 time samples without resorting to any historical TSS results and trend. Accordingly, the regions without this prediction capability are marked in light yellow. It is observed that the total area of the light blue region (capable of prediction) remarkably outweighs the area of the light yellow one (incapable of prediction). We quantify the weight of the light blue region (compared with the whole region) as the prediction accuracy rate, while the weight of the light-yellow region is considered as the training error rate. In this way, we demonstrate that the proposed TSS model index exhibits a training error rate of 32.78% and a modelling prediction accuracy rate of 67.22% for the future 15 time samples (30 working hours) of the FTSE 100 stock price.

Following a similar data manipulation procedure, we obtain the results of the traditional CEFD method in Fig. 5. By moving the curve of CEFP backwards by 8 time samples, the most identical trend between 8-time lag CEFP and the FTSE 100 stock price is reported, i.e. CEFD (CEFP) exhibits a delayed response to FTSE 100 by 16 working hours.

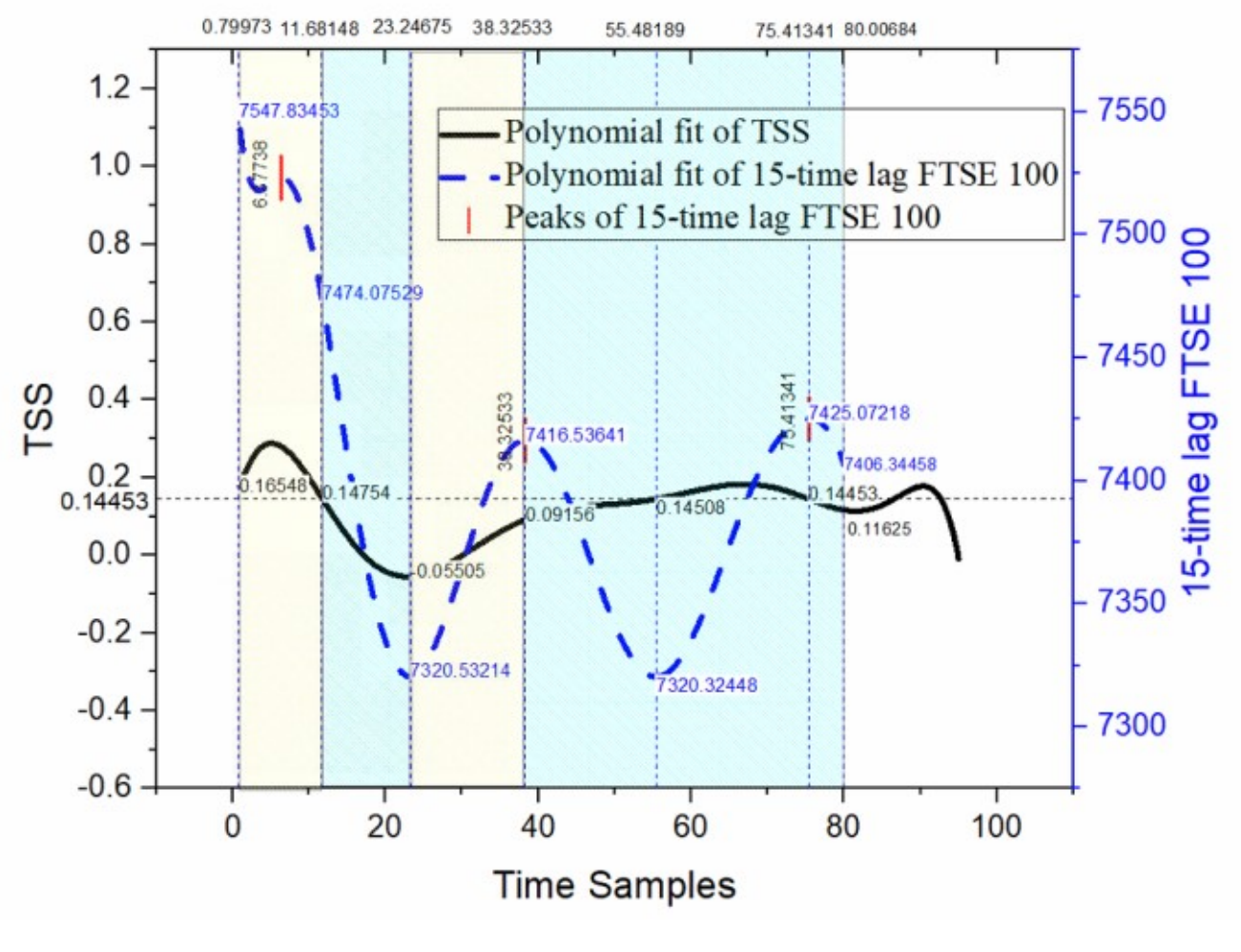

Fig. 4 Polynomial fit results of TSS and 15-time lag FTSE 100, with baseline denoted and the capable predicting areas marked in light blue.

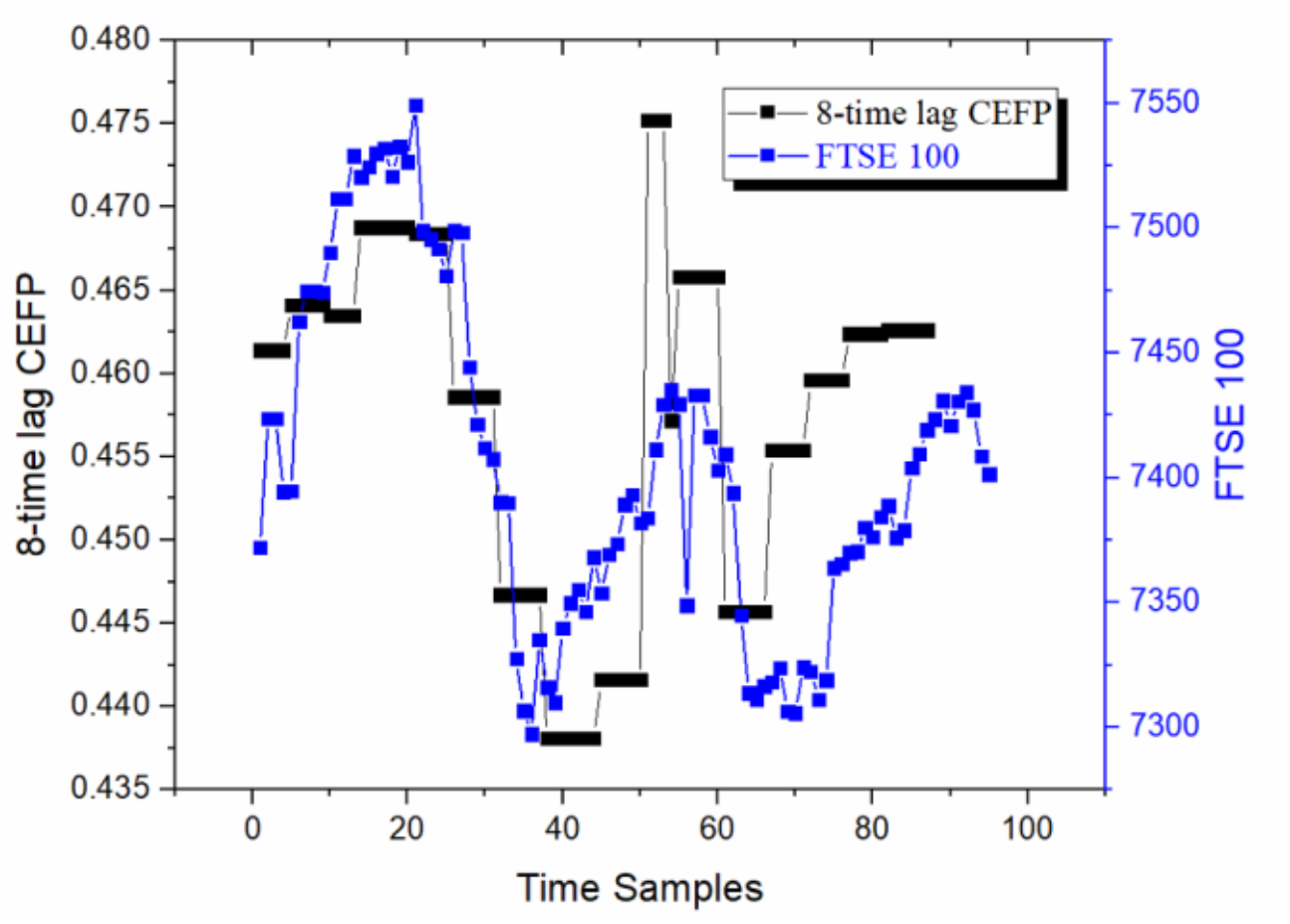

Fig. 5 Results of 8-time lag CEFP and FTSE 100.

By polynomial fitting the 8-time lag CEFP and FTSE 100 versus time, we obtain R-square of 0.9 and 0.7 respectively. With F-statistics significantly different from zero, and p value as small as zero, both polynomial fittings are effective, and we could argue that the real-time CEFP has little prediction capability for the stock price, as its current variation matches with the trend of the delayed stock market (by 16 working hours). Instead, FTSE 100 seems to exhibit a strong prediction function for 8-time lag CEFP.

In a summary of the regression model created, TSS exhibits a strong predicting functionality of the 15-time lag FTSE 100 (30 working hours in advance), with overall prediction accuracy in excess of 67.22%. Furthermore, according to the classification model used to train the above datasets and test the TSS's prediction capability by means of logistic regression and linear discriminant analysis, TSS's prediction for a future upward trend of FTSE 100 achieves an accuracy of 97.87%, while only 5.13% of accuracy is obtained for a downward market prediction.

## *B. Corpus Analysis Results*

Corpus results obtained from the TSS generating process are presented here. In a stock-related tweets keyword analysis, "Trump" turns up the biggest (highest frequently appearing) word, i.e. 78 times among 101 time samples, as shown in Fig. 6 (a), while in Fig. 6 (b), the word "Elections" appears 7 times during the British election. A decreasing sentiment score is observed during the British election period in June 2017, followed by a downward stock trend with little time lag, which illustrates the socio-political impact on the variation of investor sentimental expectation to the market and the stock price. Given the political situation fluctuates, there accompanies a downward trend or bear market of stock price, as agreed with the TSS prediction we obtained.

We observe that the prominent keyword appeared in the cloud could predict the stock trend with no time lag in response to the signal of fluctuations in the stock price. As illustrated in the afternoon of 16th Aug., when the stock market was going to close, the keyword we collected as shown in Fig. 7 (a) exhibits quite big characters of the words "buy", "red", and "popular", which is a relatively positive sentiment signal of the market. Then in the morning of 17th August, stock price indicator turned bull market in a quite short period after an 18-day sustaining-looming condition, which agrees with our corpus analysis results obtained one day in advance. For another case in Fig. 7 (b), before the stock market turned red at 10:00 on 24th August, signals of turning-down words like "alert", "signal", "short" (means "sell the stock" in financial trading market) have been reported in the twitter sentiment corpus at 8:00 on 24th Aug.

It is also worth noting that the obtained corpus analysis results could well explain our TSS prediction accuracy results being found much higher for an upward stock price trend than that of a downward trend. As shown in Fig. 8 (a), the Twitter sentiment keywords corpus shows quite a few big keywords and seemingly hot discussions on the market trend with more high-frequency words showed. Conversely for a looming stock market, the keywords corpus as reported in Fig. 8 (b) exhibits much less high-frequency keywords. Such phenomenon in a financial market indicates that investors psychologically go after gain and avoid harm, as evidenced by twittering more about bull market in the social networks while keeping away from the topic of a looming market, as a consequence, leaving more clues for a bull market sentiment analysis prediction, while less clues for a looming market. This accounts for the 92% difference in the TSS prediction accuracy for a bull (upward) market versus a looming (downward) market.

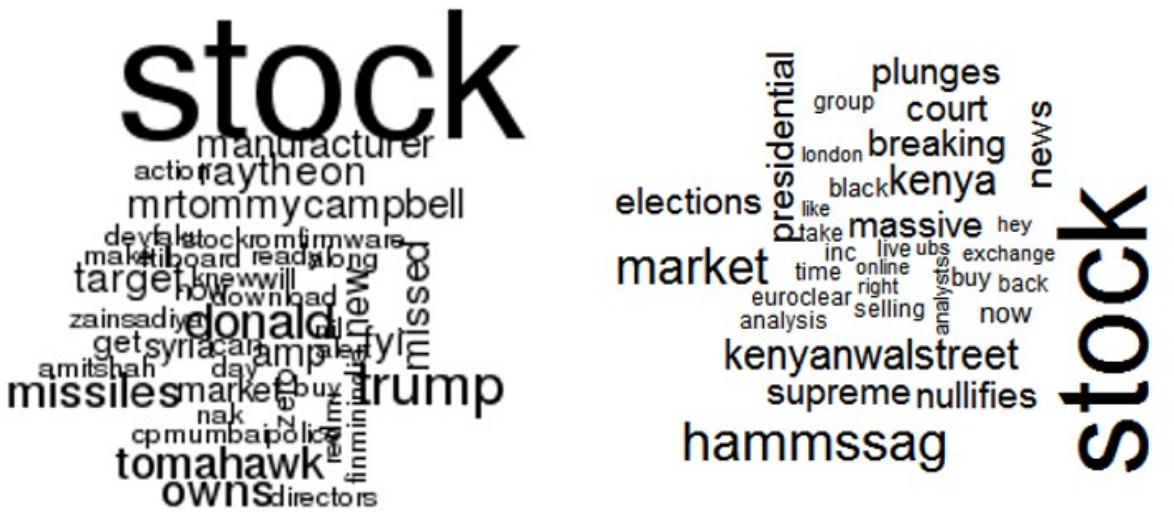

(a) (b)

Fig. 6 Corpus analysis results including high-frequency political key words (a) "Trump" and (b) "Elections".

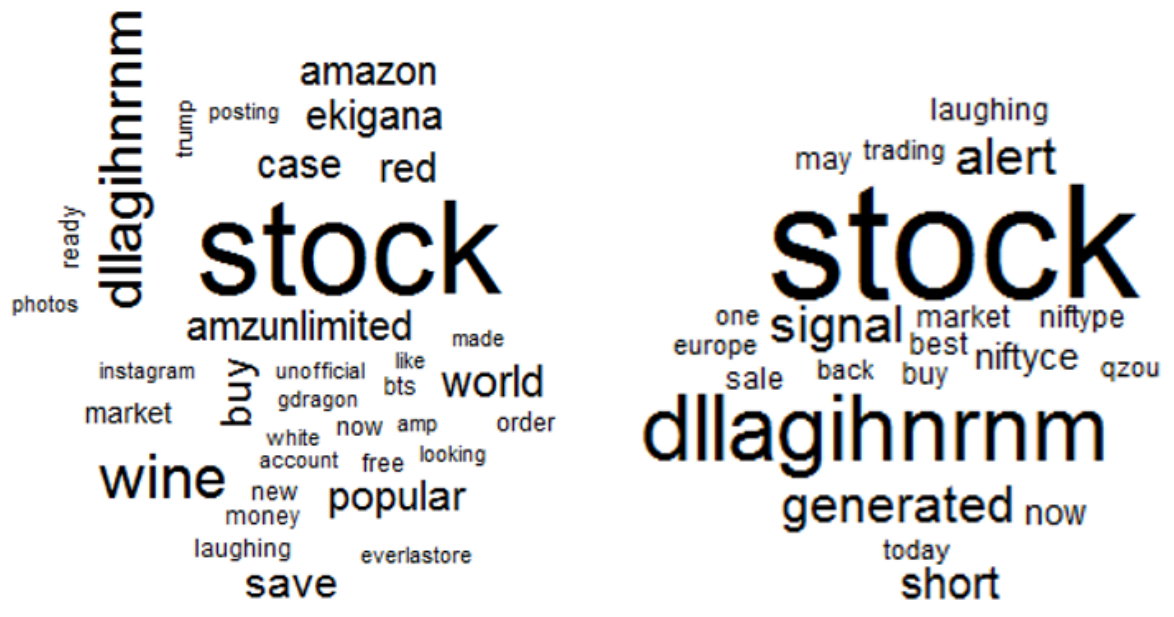

(a) (b)

Fig. 7 Corpus analysis results for (a) 16th August (stock turning green), and (b) 24th August (stock turning red).

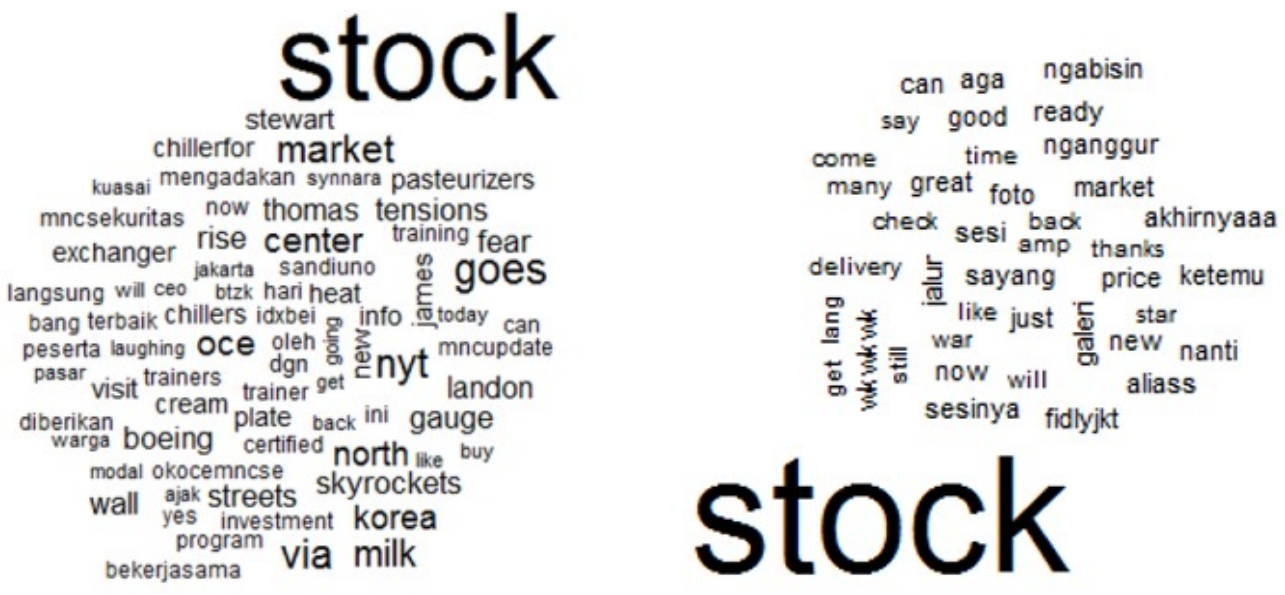

(a) (b)

Fig. 8 Corpus analysis results for (a) hot discussions in an upward stock market, and (b) cold discussions in a downward stock market.

### *C. Potential Applications and Limitations of the Approach*

The testing results above demonstrate that media sentiment score is more effective than traditional economic indicators in projecting investors' psychological transaction intention in the market, thus predicting the future market trend more accurately. A wider range of applications could benefit from this research.

Firstly, as financial market indicators are also barometers of national economics, economists can thus more accurately predict the financial market trend by monitoring TSS-based investor sentiment analysis and inform policy makers of government in the event of a financial market disruption. Secondly, the proposed model could serve as a guideline for investors and traders to make an optimal transaction judgement in advance, e.g. arbitrage transaction opportunities, investment assurance, and put-off transactions to compensate for the potential losses in the deposit duration portfolios. Thirdly, TSS can also open up more financial derivatives traded on the internet.

A possible limitation of this case study lies in the limited time series samples over a limited data tracking period. Secondly, the prediction accuracy characterizing the TSS performance depends on the baseline chosen during data processing. With an elevated baseline level, accuracy of the TSS prediction may increase or decrease, depending on the number of peaks exhibited by the fluctuated stock price and the time rate of fluctuations (i.e. slopes of the time-dependent FTSE 100 curve). This provides a future research opportunity based on the proposed TSS model with baseline correlation. One solution is by brute force attempts continuously varying the baseline, comparing the prediction accuracy accordingly, and finally deriving an optimum baseline scenario with a maximum estimation accuracy. More intelligent ways could be attempted by machine learning algorithms.

Another concern that merits discussion is the index chosen for the stock price, i.e. FTSE 100 to represent the financial market condition. Although most big enterprises covered in the FTSE 100 barometer exhibit a similar tendency with the whole market trend, we should not underestimate many small and medium companies whose share price could vary independently on the whole market, in which case that the use of the FTSE 100 would fail to estimate actual transactions. Arguably, the proposed TSS methodology can be transferrable into any other specific stock markets by modifying the collection of tweets tailored for specific market conditions with geological and demographic information factored in.

## V. CONCLUSION

This work contributes a Twitter sentiment analysis model to inform fast decision making in FTSE 100 stock market with a decent prediction accuracy. The model is based on a real-time Twitter sentiment score (TSS) with a baseline correlation that decouples the prediction model from historical TSS data trend, hence an improved efficiency as compared with up-to-date documentations. By this method in this case study, TSS is verified to exhibit predicting functionality of the future market trend in advance by 30 working hours, with an overall accuracy of 67.22% under 9th order polynomial regression fit with a baseline correlation. Specifically, TSS's prediction accuracy for an upward market trend outperforms that for a downward market, as evidenced in the accuracy of 97.87% obtained by logistic regression and linear discriminant analysis for an upward market, as compared with that in a downward market trend, in which the accuracy is only 5.13% with the same prediction model. In the calculation process of TSS using R, a keyword corpus analysis is performed to better understand the socio-cultural interactions that lead to the variation of investor sentiment and market judgement. It is envisaged that the proposed TSS model with a baseline correlation could be applied to inform fast decision-making in financial market prediction especially for a market on the rise.

## Acknowledgment

The authors would like to acknowledge Dr. Gabriele Salciute-Civiliene for the support on the research.

## References

[1] J. Bollen, H. Mao, and X. Zeng, "Twitter mood predicts the stock market," J. Comput. Science, 2011, vol. 2, pp.1-8.

[2] M. Zweig, "An investor expectations stock price predictive model using closed-end fund premiums," The Journal of Finance, 1973, vol.28, pp.67-78.

[3] C. Lee, A. Shleifer, and R. Thaler, "Investor sentiment and the closed - end fund puzzle," The Journal of Finance, 1991, vol.46, pp.75-109.

[4] C. Lee, A. Shleifer, and R. Thaler, "Anomalies: closed-end mutual funds," The Journal of Economic Perspectives, 1990, pp.153-164.

[5] N. Chopra, C. Lee, A. Shleifer, and R. Thaler, "Yes, discounts on closed-end funds are a sentiment index," The Journal of Finance, 1993, vol. 48, pp.801-808.

[6] G. Gemmill, and D. Thomas, "Noise trading, costly arbitrage, and asset prices: evidence from closed-end funds," The Journal of Finance, 2002, vol.57, pp.2571-2594.

[7] E. Kontopoulos, C. Berberidis, T. Dergiades, and N. Bassiliades, "Ontology-based sentiment analysis of twitter posts," Expert systems with applications, 2013, vol.40, pp.4065-4074.

[8] J. Bollen, A. Pepe, and H. Mao, "Modeling public mood and emotion: Twitter sentiment and socio-economic phenomena," ICWSM 2011, pp.450-453.

[9] B. Pang, L. Lee, and S. Vaithyanathan, "Sentiment classification using machine learning techniques," EMNLP 2002, pp. 79-86.

[10] B. Pang, and L. Lee, "Opinion mining and sentiment analysis," Foundations and Trends® in Information Retrieval, 2008, vol.2, pp.1-135.

[11] M. Taboada, J. Brooke, M. Tofiloski, K. Voll, and M. Stede, " Lexicon-based methods for sentiment analysis," Computational Linguistics, 2011, vol.37, pp.267-307.